\documentclass[aps,pre,subeqns.floatfix,amsmath,twocolumn]{revtex4}
\usepackage[pdftex]{graphicx}
\usepackage{amsmath}
\usepackage{amsfonts}
\usepackage{graphicx}
\usepackage{hyperref}
\usepackage{comment}
\usepackage{mathtools}
\usepackage{verbatim}
\usepackage[outdir=./]{epstopdf}
\usepackage{subfigure}
\usepackage{hyperref}
\usepackage{latexsym}
\usepackage{dcolumn}
\usepackage{epsf}
\usepackage{float}

\usepackage{color}
\usepackage[utf8]{inputenc}

\begin{document}

\title{Impact of phase lag on synchronization in frustrated Kuramoto model with higher-order interactions}

\author{Sangita Dutta$^1$}
\email{sangitaduttaprl@gmail.com}
\author{Abhijit Mondal$^1$}
\author{Prosenjit Kundu$^2$}
\email{jitprosen.math@gmail.com}
\author{Pitambar Khanra$^3$}
\email{pitambar.khanra89@gmail.com}
\author{Pinaki Pal$^1$}
\email{pinaki.pal@maths.nitdgp.ac.in}
\author{Chittaranjan Hens$^4$}
\affiliation{$^1$Department of Mathematics, National Institute of Technology, Durgapur~713209, India}
\affiliation{$^2$Dhirubhai Ambani Institute of Information and Communication Technology, Gandhinagar, Gujarat, 382007, India}
\affiliation{$^3$Department of Mathematics,  State University of New York at Buffalo, Buffalo, USA}
\affiliation{$^4$Center for Computational Natural Science and Bioinformatics, International Institute of Informational Technology, Gachibowli, Hyderabad 500032, India}
\begin{abstract}
The study of first order transition (explosive synchronization) in an ensemble (network) of coupled oscillators has been the topic of paramount interest among the researchers for more than one decade. Several frameworks have been proposed to induce explosive synchronization in a network and it has been reported that phase frustration in a network usually suppresses first order transition in the presence of pairwise interactions among the oscillators.  However, on the contrary, by considering networks of phase frustrated coupled oscillators in the presence of higher order interactions (upto $2$-simplexes) we show here under certain conditions, phase frustration can promote explosive synchronization in a network. A reduced order model of the network in the thermodynamic limit is derived using the Ott-Antonsen ansatz to explain this surprising result. Analytical treatment of the reduced order model including bifurcation analysis explains the apparent counter intuitive result quite clearly. 

\end{abstract}
\maketitle
\section{Introduction}

Synchronization \cite{Strogatz_synchronization_book, Pikovsky_synchronization_book, kuramoto1984chemical} is a captivating phenomenon observed in natural and artificial systems  \cite{wang2017explosive,kumar2015experimental,lee2018functional,blasius2000chaos,pantaleone2002synchronization,grzybowski2016synchronization,Kuramoto_chemical_book,buck1988synchronous,yamaguchi2003synchronization,wiesenfeld1998frequency,dorfler2014synchronization}. A large set of coupled oscillators undergoes a continuous  synchronization transition if the interaction strength among themselves is gradually increased. To describe such transition, a simple  and analytically tractable phase model was developed by Yoshiki Kuramoto~\cite{Kuramoto_chemical_book}, in which each oscillator  has its own intrinsic frequency ($\omega_i$), and cross-talks to each other via a periodic coupling function representing pairwise interactions. The equation of motion of each oscillator under all-to-all coupling configuration is captured by
\begin{eqnarray}
\label{clasicKM}
\dot{\theta_i}=\omega_i+K \sum_{j=1}^{N}\sin(\theta_j-\theta_i), \hspace*{.4cm} i=1,2,\dots,N,
\end{eqnarray}
where $K$ is the coupling strength.  To measure the level of synchronization, the order parameter $r$ is defined as 
\begin{eqnarray}
\label{clasicKM}
re^{i\psi}= \sum_{j=1}^{N}e^{i\theta_j}.
\end{eqnarray}
The classical Kuramoto model (KM) described above shows second order (continuous) transition to  global synchronization ($r\rightarrow 1$) state as the coupling strength is gradually  increased from a small value. All the oscillators rotate with the mean of the natural frequencies in a higher coupling strength.
However, different scenario may appear if a phase-lag or frustration parameter ($\beta$) \cite{sakaguchi1986soluble, omel2012nonuniversal,  kundu2017transition, kundu2019synchronization, kundu2018perfect, brede2016frustration,lohe2015synchronization,english2016emergence,wiesenfeld1998frequency,lohe2010quantum,dorfler2012synchronization} is introduced in the pairwise coupling function ($\sin(\theta_j-\theta_i-\beta)$). This phase-lag parameter   shifts the system's synchronous frequency from the mean natural frequency. Thus, $\beta$ can be used as a control parameter that can tune the mean frequency to a desired one~\cite{lohe2015synchronization}. The model was first introduced by  H. Sakaguchi jointly with Y. Kuramoto and presently known as Sakaguchi-Kuramoto (SK) model. In SK model, the phase-lag $\beta$ generally opposes the coupled system to reach to the global synchronous state and consequently, higher coupling strength compared to KM is required for achieving a desired level of synchronization.  Also, for a particular choice of frequencies, the synchronization may deviate from the universal continuous transition route~\cite{omel2012nonuniversal,omel2013bifurcations,omel2016there}. Further it has been shown, that in a complex network setup, the SK model cannot reach a global synchronous state (called as {\it erosion} of synchronization)~\cite{skardal2015erosion}. Even over a critical $\beta$, where the natural frequencies are correlated with degree, the hysteresis or explosive width can be entirely vanished, and the level of synchronization would be significantly reduced~\cite{kundu2017transition}. To overcome such situations, the design of suitable frequencies~\cite{brede2016frustration,kundu2018perfect,skardal2014optimal}, usage of multiple layers of networks~\cite{khanra2018explosive, kundu2020optimizing,khanra_chaossoli2021}, and time-dependent coupling functions~\cite{khanra2020amplification} have been proposed. Note that, models of Josephson junction arrays~\cite{wiesenfeld1998frequency}, power network systems~\cite{dorfler2012synchronization}, dynamics of mechanical rotors~\cite{mertens2011synchronization}, quantum networks~\cite{lohe2010quantum} all can be captured by such phase-lag oscillators under the paradigm of pairwise interactions. 
 \begin{figure}
     \includegraphics[width=0.25\textwidth]{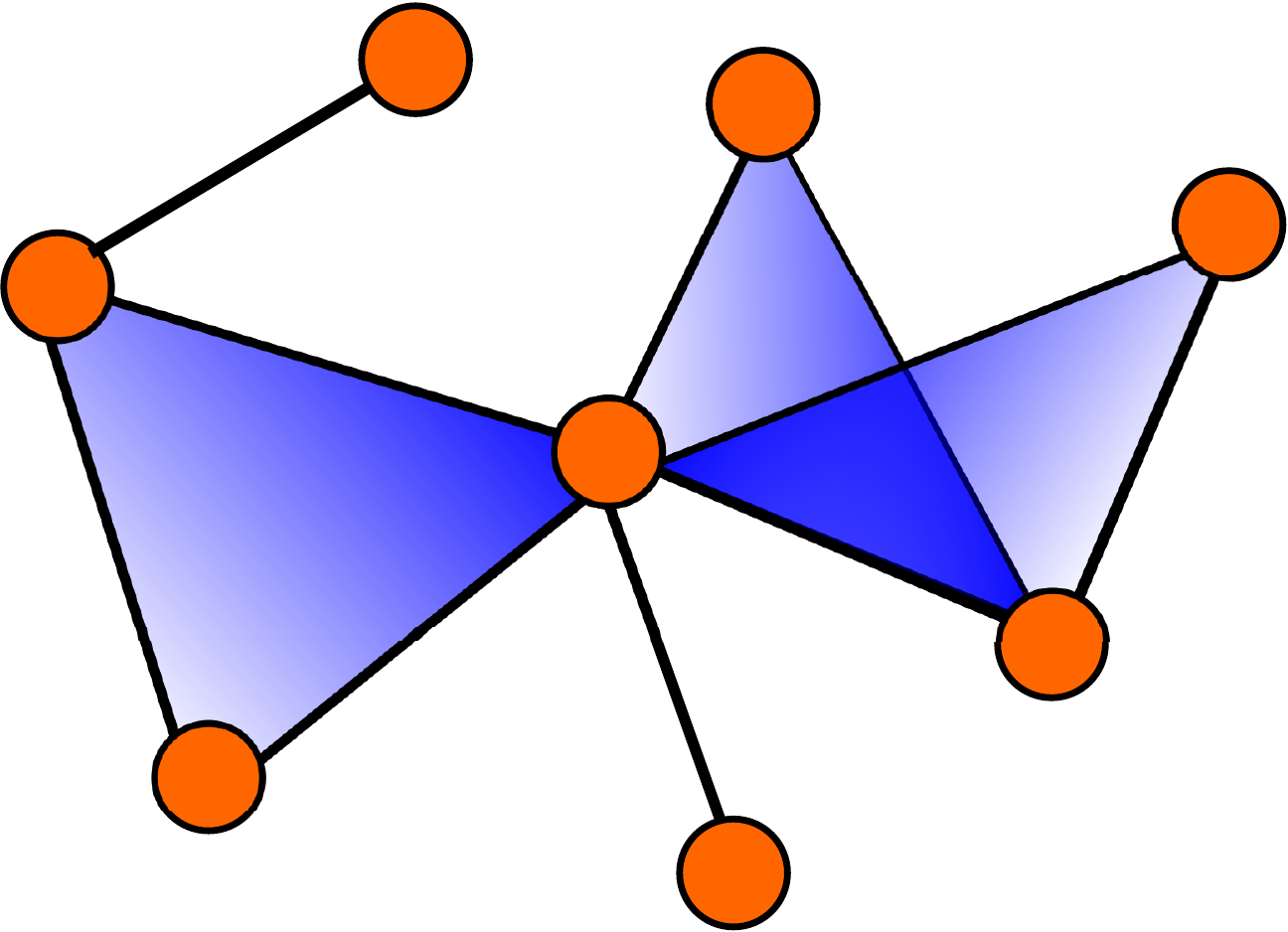}
\caption{Schematic diagram of a simplicial complex consisting of $0$-simplex (circles), $1$-simplex (edges) and $2$-simplex (triangles).}
     \label{fig:simplex}
 \end{figure}

\begin{figure*}
         \includegraphics[width=1\textwidth]{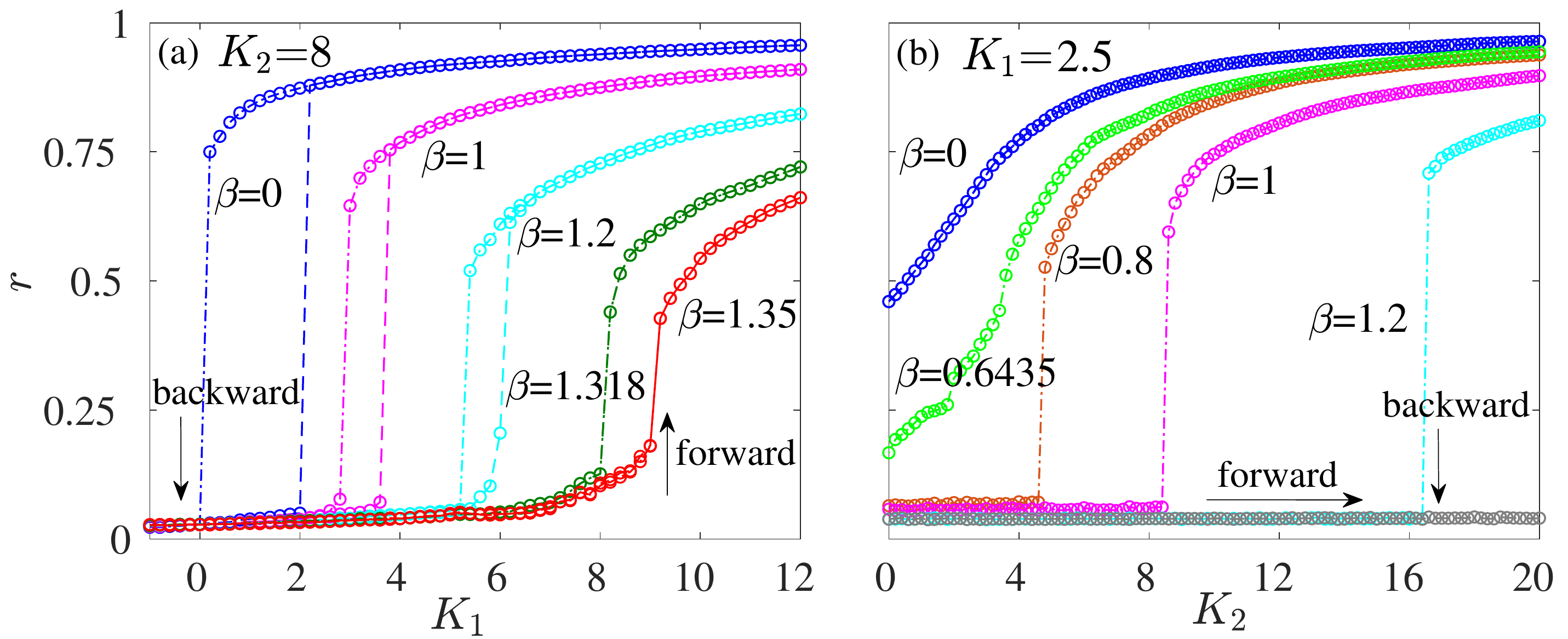}
     \caption{Synchronization profile for the all-to-all connected network with different phase-lag values as obtained from the numerical simulation. Forward and backward transitions are indicated by up and down arrows respectively. (a) Numerically computed $r$ is shown as a function of pairwise coupling $K_1$ for fixed triadic coupling $K_2=8$.  As $\beta$ changes from $0$ to higher values, the transition gradually becomes  continuous from discontinuous. (b) For fixed $K_1=2.5$, $r$ is plotted as a function of $K_2$ for various $\beta$. Larger values of lag induces bistable states in the system.}
     \label{fig1}
     \end{figure*}
However, in recent studies the importance of incorporating higher order interactions (HOI) along with its pairwise counterpart has been emphasized~\cite{majhi2022dynamics,battiston2020networks,alvarez2021evolutionary,iacopini2019simplicial,martignon1995detecting,yu2011higher,giusti2016two,reimann2017cliques,tanaka2011multistable}. Specially the works on neuroscience \cite{yu2011higher,giusti2016two,reimann2017cliques,sizemore2018cliques}, ecology \cite{chatterjee2022controlling,grilli2017higher,bairey2016high,kareiva1994special} and physics \cite{ghasemi2022higher} have highlighted the crucial role of such higher order interactions ( e.g. in a collaboration network multi authors collaborate \cite{vasilyeva2021multilayer}) in addition to pairwise one. Simplicial complexes \cite{salnikov2018simplicial,costa2016random} can successfully encode these higher order structures.  A $n$-simplex is formed by $(n+1)$ interacting units, consisting all the $d$ ($<n$) simplexes. For example, $1$-simplex denotes the pairwise interactions, $2$-simplex denotes the three way interactions including pairwise one and so on. These simplices adhere to one another along their sides, and form a simplicial complex. FIG.~\ref{fig:simplex} represents schematic diagram of a simplicial complex with higher order interactions. Here nodes (circles) are connected with edges and triangles (blue). In case of triangles, pairwise connections between the three nodes are also taken into account. Till now some of the dynamical behaviors induced by the HOI have been explored including multistability \cite{skardal2019abrupt,xu2020bifurcation}, chimera states \cite{zhang2021unified}, chaos \cite{bick2016chaos} etc. Also it has been shown that the presence of HOI along with pairwise interactions can produce high level of synchronization in weaker coupling strength \cite{skardal2021higher} and may lead to the first order or discontinuous transition to synchronization \cite{skardal2019abrupt}.

Recently, Skardal {\it et al}.\ has reported in \cite{skardal2020higher} that the HOI in a simplicial complex can exhibit abrupt transition to synchrony forming hysteresis, without any degree frequency correlation. These synchronized states are stable even in negative pairwise coupling for the larger values of HOI coupling strengths. They analytically derived the evolution equation for the order parameter to observe the macroscopic dynamics in all-to-all network and analyze the bifurcations happen in the considered parameter space. In \cite{rajwani2023tiered}, it has been demonstrated that adaptation to higher order coupling leads to tired synchronization transition in addition with second order transition.

In this letter, we consider SK model with higher order interactions (upto $2$-simplex) under the all-to-all coupling configuration and observe a significantly different effect of phase frustration compared to the ones observed in the presence of pairwise interactions only. It is observed that in the presence of HOI,  the increment of phase frustration eliminates first order transition in the system as the pairwise coupling is varied for a fixed HOI coupling strength. On the other hand, for a fixed pairwise coupling strength as the HOI coupling strength is varied, first-order transition is induced in the system even for very high values of $\beta$.

Along with numerical simulations, here we present an analytical framework using Ott-Antonsen anastz~\cite{ott2008low} to derive a reduced order model of the system. The reduced order model captures the dynamics of the entire system quite accurately in the parameter regime of our interest and explains the origin of different kinds of transitions in the system. It is observed that the first order transition is associated with the subcritical pitchfork bifurcation, while, second order transition is associated with supercritical pitchfork bifurcation. In other words, the bifurcation analysis of the reduced order equation provides comprehensive understanding of the incoherent, coherent and bistable states observed in numerical simulations.

\section{Model Description}    
The dynamics of a phase frustrated undirected simplicial complex of $N$ number of nodes with global connectivity is governed by the following system of equations
\begin{eqnarray}
\label{model}
\dot{\theta_i}=\omega_i&+&\frac{K_1}{N} \sum_{j=1}^{N}\sin(\theta_j-\theta_i-\beta)      \nonumber\\ 
&+&\frac{K_2}{N^2} \sum_{j=1}^N \sum_{l=1}^N \sin(2\theta_j-\theta_l-\theta_i-\beta),  \label{SKHOI}\\ 
&&\hspace*{3.5cm} i=1,2,\dots,N.  \nonumber
\end{eqnarray}
This is the generalization of classical Sakaguchi-Kuramoto model to a model consisting of higher order interactions in addition to pairwise ones.
Here $\theta_i$ represents the phase  and $\omega_i$ is the natural frequency of the $i$th oscillator, $\omega$ is drawn from a Lorentzian distribution with density function $g(\omega)=\frac{\Delta}{\pi[\Delta^2+(\omega-\omega_0^2)]}$ with mean $\omega_0$. $\beta$ denotes the uniform phase lag of the system.  $K_1$ and $K_2$ are the coupling strengths of pairwise interactions and HOI respectively. 

Two complex order parameters associated with $1$-simplex and $2$-simplex are defined by 
\begin{eqnarray}
\label{order_parameter}
z=re^{i\psi_1}=\frac{1}{N}\sum_{j=1}^{N}e^{i\theta_j},  \hspace*{.5cm} z_2=r_2e^{i\psi_2}=\frac{1}{N}\sum_{j=1}^{N}e^{2i\theta_j}
\end{eqnarray}
where the amplitudes $r=\vert z \vert$ and $r_2=\vert z_2 \vert$ of the order parameters $z$ and $z_2$ respectively measures the level of synchronization, $\psi_1$ and $\psi_2$ are the respective arguments, representing the average phases of the oscillators.
\section{Numerical observation}\label{sec:obser}
To find the effect of the phase frustration on SK model with HOI,  we simulate the equation (\ref{SKHOI}) with network size $N = 10^3$. For the simulation, $4$-th order Runge-Kutta method with time step $\delta t=0.01$ is used. First we check the effect of phase lag $\beta$ with fixed $K_2$ ($= 8$). Here the values of $\beta$ are varied in the range of  $[0,\frac{\pi}{2})$. Initially the phases of the oscillators are spread uniformly at random  around a circle which means that the initial condition are chosen randomly from $[-\pi,\pi]$. The natural frequencies are drawn from a Lorenzian distribution with mean $0$ and half width $\Delta =1$. To determine the nature of transitions to synchronization we continue the simulation of  the system (\ref{SKHOI}) by varying the relevant parameter both in forward and backward directions. 
We start the simulation with $K_1=-1$ and gradually increase the pairwise coupling strength $K_1$ to achieve the synchronized state for fixed $K_2 = 8$. For backward transition we start from the synchronized state achieved with the simulation for forward transition and gradually decrease $K_1$ until we reach  to $K_1=-1$. Note that, during forward and backward continuation, the final state of the previous simulation is taken as the initial condition for the present simulation. FIG. \ref{fig1}(a) shows the variation of $r$ with $K_1$ for different values of $\beta$ namely, $\beta=0$ (blue), $\beta=1$ (magenta), $\beta=1.2$ (cyan), $\beta=1.318$ (green), and $\beta=1.35$ (red) for both forward and backward continuations. The figure shows that the system exhibits explosive synchronization transition for smaller values of $\beta$ and the transition becomes continuous as $\beta$ is gradually increased  from $0$ to $\frac{\pi}{2}$. Also, it is evident from FIG. \ref{fig1}(a) that the hysteresis width for the synchronization transition decreases with the increase of $\beta$ and the transition become second order or continuous for $\beta > 1.318$. 

Next, to understand the impact of HOI coupling strength $K_2$ on the synchronization transition, we simulate the system (\ref{SKHOI}) by varying $K_2$ with fixed $K_1 = 2.5$ for different $\beta$. The variation of the synchronization order parameter $r$ with  respect to $K_2$ (both for forward and backward continuation, and for different $\beta$) is presented in the FIG.~\ref{fig1}(b).  From the figure one can observe for a low $\beta$ value ( $\beta =0$ (blue) and $0.6435$ (bright green)), $r$ follows a continuous transition, i.e.  forward and backward continuation data follow the same path. On the other hand, for $\beta > 0.6435$, say for $\beta = 0.8$ (brown), $1$ (magenta) and $1.2$ (cyan) discontinuous (first order) transition to synchronization is observed. Therefore, from the simulation result, it is clearly observed that in this case with the increase of $\beta$ the backward transition points move forward but explosive synchronization is not suppressed by higher phase frustration of the system.   This unexpected result seems to be a unique feature of HOI. Here the effect is opposite to the one that is observed in presence of pairwise interactions only, where phase frustration is known to suppress the discontinuous transition to synchronization~\cite{kundu2017transition}. To develop a deeper understanding about the above surprising result we proceed to derive a reduced order model of the system~(\ref{SKHOI}) using Ott-Antonsen ansatz~\cite{ott2008low}. The details are described in the next section.

\section{A reduced order model}
As mentioned earlier, in this section we derive a reduced order model to explain the numerically observed transition to synchronization in the coupled system under consideration.  First to simplify the onward calculation  we write the equation (\ref{model}) with the help of equation (\ref{order_parameter}) as follows
\begin{equation}
\label{reduced_model}
\dot{\theta_i}=\omega_i+\frac{1}{2i}\left[e^{-i(\theta_i+\beta)}H-e^{i(\theta_i+\beta)}\bar{H}\right],
\end{equation}
where $H=K_1z+K_2z_2\bar{z}$ and overbar denotes complex conjugate. In the thermodynamic limit, we assume that the density of the oscillators with phase $\theta$, frequency $\omega$ at time $t$ is given by $f(\theta,\omega,t)$. Since the natural frequency is drawn from a distribution $g(\omega)$ the function $f$ can be expanded in a Fourier series as
\begin{equation}
f=\frac{g(\omega)}{2\pi}\left[1+\sum_{n=1}^{\infty}[f_n(\omega,t)e^{in\theta}+c.c.]\right], \label{ott-FS}
\end{equation}
here $c.c.$ denotes the complex conjugate and $f_n(\omega,t)$ is the co-efficient of the $n$-th term of the series. Now following Ott-Antonsen ansatz~\cite{ott2008low} we take $f_n=\alpha^n$ for some analytic function $\alpha$.
Since the oscillators are conserved in a network, the density function must satisfy the continuity equation 

\begin{equation}
\label{continuty_eq}
\frac{\partial f}{\partial t}+\frac{\partial}{\partial \theta}{(fv)}=0,
\end{equation}
where $v=\frac{d\theta}{dt}$ is the velocity given by equation (\ref{reduced_model}).
Replacing $f$ by its Fourier series representation (\ref{ott-FS}) in the continuity equation we obtain 
\begin{equation}
\label{eq_alpha}
\dot{\alpha}+i\alpha\omega+\frac{1}{2}\left[H\alpha^2e^{-i\beta}-\bar{H}e^{i\beta}\right]=0.
\end{equation} 
Now, in continuous form, the complex order parameter can be written as 
\begin{eqnarray}
\bar{z} &=&\int d\omega \int \bar{f}(\theta,\omega,t)e^{-i\theta}d\theta \nonumber\\
&=& \int d\omega \int_0^{2\pi} \frac{g(\omega)}{2\pi}[1+\sum_{n=1}^{\infty}[\bar{\alpha}^n(\omega,t)e^{-in\theta} \nonumber\\
&&\hspace*{2.5cm} +\alpha^n(\omega,t)e^{in\theta}]]e^{-i\theta} d\theta \nonumber \\
&=& \int \alpha(\omega ,t)g(\omega)d\omega.
\end{eqnarray}
Integrating this in the lower half of the complex plane using the Cauchy's integral theorem we get, $\bar{z}=\alpha(\omega_0-i\Delta,t)$. Similarly we can derive $\bar{z}_2=\alpha^2(\omega_0-i\Delta,t)={\bar{z}}^2$. 

At $\omega=\omega_0-i\Delta$, equation (\ref{eq_alpha}) leads to
\begin{eqnarray*}
\dot{\bar{z}}=-i\bar{z}(\omega_0-i\Delta)+\frac{1}{2}[(K_1\bar{z}+K_2\bar{z_2}z)e^{i\beta}   \\
-{\bar{z}}^2(K_1 z+K_2 z_2\bar{z})e^{-i\beta}].
\end{eqnarray*}
Taking complex conjugate on both sides we get
\begin{eqnarray}
\label{eq_order_parameter_z}
\dot{z}=iz\omega_0-z\Delta+\frac{1}{2}[(K_1z+K_2z^2\bar{z})e^{-i\beta} \nonumber \\
-z^2(K_1\bar{z}+K_2{\bar{z}}^2z)e^{i\beta}].
\end{eqnarray}
Putting $z=re^{i\psi}$ and comparing real and imaginary parts on both sides of equation (\ref{eq_order_parameter_z}) we obtain the following reduced order model for the system (\ref{SKHOI})
\begin{eqnarray}
\dot{r}&=& -\Delta r+\frac{1}{2}[(K_1 r+K_2 r^3)\cos\beta\nonumber\\
&-&(K_1 r+K_2 r^3)r^2\cos\beta],\label{rom}
\label{eq_r}
\end{eqnarray}
\begin{eqnarray}
\dot{\psi}= \omega_0+\frac{1}{2}[-(K_1+K_2r^2)\sin\beta \nonumber \\
-(K_1+K_2r^2)r^2\sin\beta].  
\label{eq_psi}
\end{eqnarray}
Now using the above reduced order model  we proceed to explain the numerically observed  collective dynamics of the considered system.
\section{Analytical VS. Numerical results}
\begin{figure}
    \includegraphics[width=0.48\textwidth]{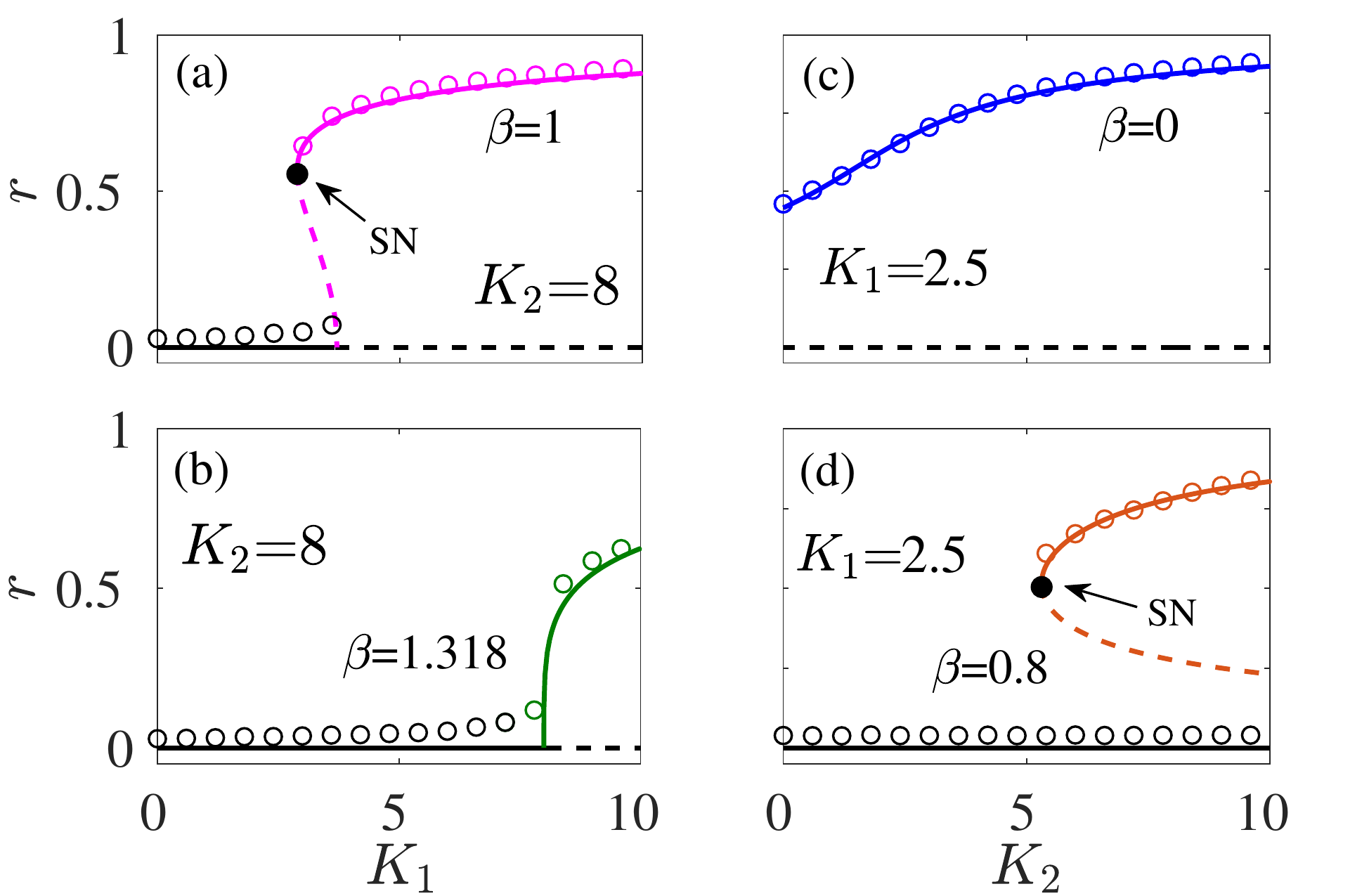}
    \caption{Bifurcation diagrams prepared using the reduced order model. First column [(a)-(b)]: $r$ as a function of $K_1$ for $K_2 = 8$ with $\beta=1$ and $1.318$; Second column [(c)-(d)]: the variation of $r$ with $K_2$ for $K_1 = 2.5$ with $\beta=0$ and $0.8$ respectively. Solid and dashed lines are obtained from the ROM, and respectively represent the stable and unstable equilibria. The filled black circles show the saddle-node (SN) bifurcation points. The colored circles show the values of $r$ corresponding to the stable solutions as obtained in the numerical simulation.}
    \label{fig:pitchfork}
\end{figure}

We perform bifurcation analysis of the reduced order model (\ref{rom}) using the software MATCONT~\cite{dhooge2003matcont} to understand the origin of the numerically observed solutions obtained from the system~(\ref{SKHOI}). First, we take $K_2 = 8$ and construct the bifurcation diagrams for $\beta = 1$ and $1.318$ by varying $K_1$. FIG.~\ref{fig:pitchfork}(a) and (b) shows the bifurcation diagrams where variation of $r$ with $K_1$ have been presented along with the stability information. For $\beta = 1$, the trivial $r = 0$ solution (solid and dashed black line) becomes unstable via a subcrcitical pitchfork bifurcation at $K_1 =3.6$.  An unstable branch originated from there moves backward (dashed magenta curve) and eventually becomes stable via a saddle-node (SN) bifurcation at $K_1 = 2.9$, and move forward (solid magenta curve) as a stable branch. Thus, in the range  $2.9 \leq K_1 \leq 3.6$ bistability induced by the subcritical pitchfork bifurcation is observed and it is responsible for the appearance of first order transition at $\beta = 1$. Further, on top of the bifurcation diagram, some data points obtained from the numerical simulation (magenta and black circles) are also plotted and a very good agreement is observed. On the other hand, for $\beta = 1.318$, $r = 0$ branch (solid and black line) becomes unstable via supercritical pitchfork bifurcation  $K_1 = 8$ and  a stable branch is originated (solid green branch). As a result, a continuous transition occurs for $\beta = 1.318$. In this case also, numerical simulation data (green and black circles) closely matches with the analytical ones.  

Next with fixed $K_1=2.5$ we construct two bifurcation diagrams for $\beta = 0$ and $0.8$ by varying $K_2$ (FIG.~\ref{fig:pitchfork}(c) and (d)). For $\beta = 0$ (solid blue curve), we observe continuous transition and the trivial $r = 0$ branch (dashed black line) is always unstable in this case. On the other hand, for $\beta = 0.8$, zero solution is always stable (solid black line). In addition to that, another pair of non-zero fixed points of the reduced system originated from a SN bifurcation point at $K_2 = 5.3$, of which one is stable (solid brown curve) and other one is unstable (dashed brown curve). As a result, bistability occurs for all $K_2 \geq 5.3.$  Here, the SN bifurcation is responsible for the discontinuous transition to synchronization. In this case also along with the analytical curve we plot the data obtained from the numerical simulation (brown and black circles) and we observe a very close agreement. It is now apparent that the reduced order model nicely mimics the dynamics of the whole system. Therefore we proceed for more detailed analysis using the model.

It is important to note here that for fixed $K_2$ as $\beta$ is increased starting from $0$, the system moves from discontinuous to continuous transition states with the variation of $K_1$. The discontinuous transition here is associated with the existence of the SN bifurcation point and it ceased to exist as the SN point meets the trivial ($r=0$) fixed point. This observation helps to determine the starting point of continuous transition. 

The equilibrium points of the reduced order model (\ref{rom}) are given by 

\begin{widetext}
\begin{equation}
r = 0~\mathrm{and}~r=\sqrt{\frac{K_2\cos\beta-K_1\cos\beta\pm\sqrt{(K_1\cos\beta+K_2\cos\beta)^2-8K_2\cos\beta}}{2K_2\cos\beta}}.\label{FP}
\end{equation}
\end{widetext}

Out of two nonzero equilibrium points given in (\ref{FP}), the one obtained with the positive sign (say $r^+$) is stable and the other one (say $r^-$) is unstable in the entire ranges of their existence in the parameter space. Now the SN bifurcation will occur for the parameter values for which $r^+ = r^-$ and it leads to the relation 
\begin{equation}
\label{saddle_point}
(K_1+K_2)^2\cos\beta=8K_2
\end{equation}
involving the parameters $K_1$, $K_2$ and $\beta$. Thus, for any two given parameters, a feasible solution of the above equation for the third one will provide the SN bifurcation point.  Now when the saddle node point merges with the fixed point $r=0$, the hysteresis will vanish and the system will follow a continuous transition path. Imposing the condition (\ref{saddle_point}) and $r=0$ in the equation (\ref{FP}) we can find the point marking the onset of continuous transition, which gives
\begin{equation}
\label{condition}
K_2\cos\beta-K_1\cos\beta=0.
\end{equation}
Solving the equation (\ref{saddle_point}) and (\ref{condition}),  the condition for the onset of continuous transition is obtained as
\begin{equation}
\label{critical_point}
    K_1=K_2=\frac{2}{\cos \beta}.
\end{equation}
Now in particular, for fixed $K_2=8$, the onset of continuous transition occurs for $K_1=8$ and $\beta = \cos^{-1}(1/4) = 1.3181.$ In the numerical simulation also, the continuous transition start to occurs exactly at this analytically predicted value. 

On the other hand, for fixed $K_1$, as $\beta$ is increased, completely opposite scenario occurs, i.e. in this case, the the system moves from continuous to discontinuous transition states with the variation of $K_2$.  The trivial fixed point $r=0$ is unstable or stable according as $\beta < ~\mathrm{or} > ~\cos^{-1}\left(\frac{2\Delta}{K_1}\right)$. In particular, for $\Delta = 1$ and $K_1=2.5$, $\cos^{-1}\left(\frac{2\Delta}{K_1}\right) = 0.6435$. Thus, for $\beta < 0.6435$, the trivial equilibrium point is unstable and it is stable for $\beta > 0.6435$.   It can be easily checked  that for $\Delta =1$, $K_1=2.5$, and $\beta < 0.6435$ only one nontrivial equilibrium point i.e. $r^+$ will exist and stable.  As a result, only second order transition to synchronization  is expected to occur in this parameter regime and it is clearly visible from the numerically obtained blue and bright green curves presented in the FIG.~\ref{fig1}(b). Whenever, $\beta > 0.6435$ along with the trivial stable fixed point, both the nontrivial fixed points $r^+$ (stable) and $r^-$ (unstable) will exist which marks the existence of bistability in the system that is responsible for first order transition. The analytical result matches very closely  with the simulation results presented in the FIG.~\ref{fig1}(b).  

\begin{figure*}
        \includegraphics[width=1\textwidth]{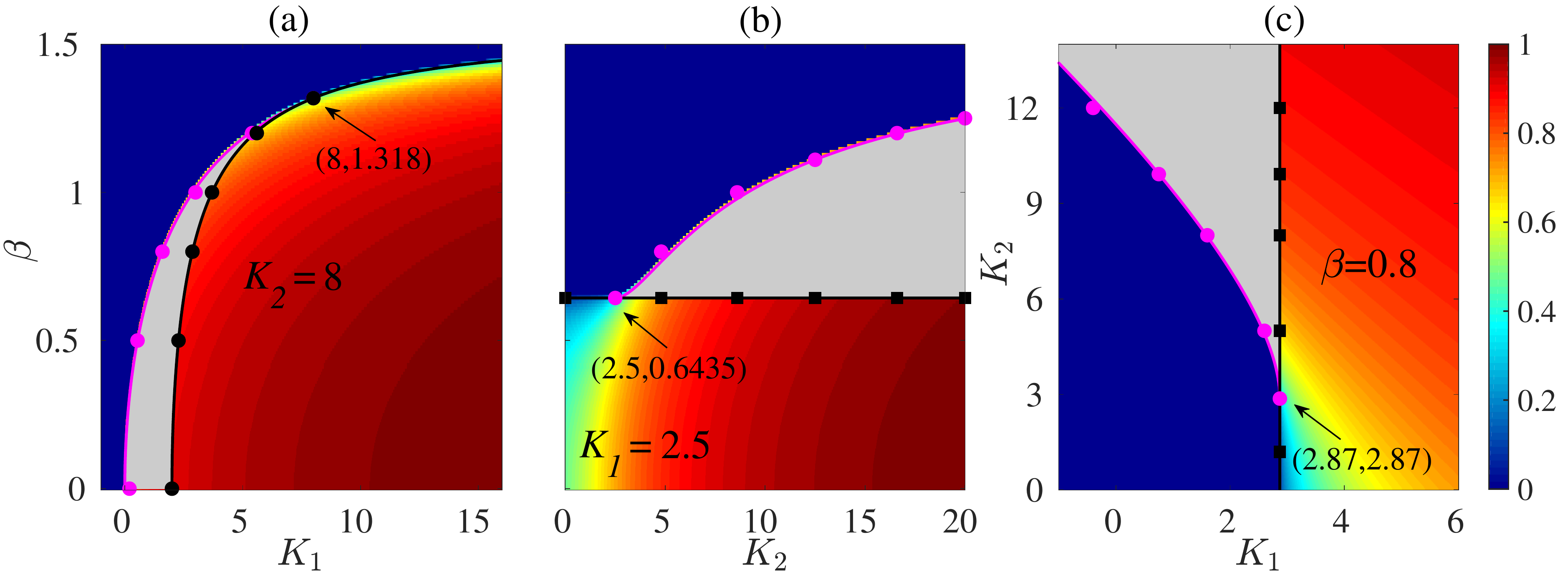}
        \caption{Two parameter stability diagrams prepared from the ROM showing the incoherent (blue), coherent (cyan to maroon) and bistable (grey) regimes.  Stability diagrams: (a) The solid magenta (backward transition or SN points) and black curves (forward transition point) delimit different synchronization regimes on the $K_1 - \beta$ plane for $K_2 = 8.$   The magenta and the black curves intersect at the point $(8,1.318)$ marking the onset of continuous transition. (b) Shows the synchronization regimes on the $K_2 - \beta$ plane for $K_1=2.5$. Below the critical lag (black line) synchronization level is high and above this value incoherent and bistable regime are separated by the saddle node curve (solid magenta curve). (c) Different regimes on the $K_1 - K_2$ plane for $\beta=0.8$.  The bistable regime starts from the intersection point $(2.87,2.87)$ of saddle node and pitchfork bifurcation curves (supercritical for $K_2<2.87$, subcritical for $K_2>2.87$) and expands with increasing $K_2$. Black and magenta dots respectively represent the corresponding transition points obtained from the numerical simulations in all the cases. The black filled squares in (b) and (c) respectively represent the numerically obtained critical $\beta$ and $K_1$ for the onset of discontinuous transitions.}
        \label{fig:stability}
    \end{figure*}

Inspired by the close agreement of the reduced order model (ROM) and the numerical simulations results, using the ROM we construct stability diagrams on different projections of the parameter space and present in the FIG.~\ref{fig:stability}. FIG.~\ref{fig:stability}(a) shows the bifurcations in the $K_1$ - $\beta$ plane for fixed $K_2 =8$.  The magenta curve indicates the backward transition points or the points of SN bifurcation (given by  equation (\ref{saddle_point}))and the black curve indicates the forward transition points (given by equation (\ref{critical_point})). The region bounded by these two curves depicts the bistable region. It is observed that the area of bistability reduces with the increase of $K_1$. Furthermore, when these two curves meet, the bistability vanishes at the critical point $(8,1.318)$ and the system exhibits synchronization via continuous paths (see FIG.~\ref{fig1}(a)) only. At these forward transition points for different lag, the system shows pitchfork bifurcation. When $\beta$ is greater than (less than) the critical value $1.318$, the  supercritical (subcritical) pitchfork bifurcation is observed. Here we use colorbar to indicate the values of $r$ calculated using the equation(\ref{FP}) considering the stable one. The blue side is for incoherent state, as the color changes from blue to maroon the value of the order parameter $r$ increases from $0$ to $1$. The bistable region is shaded by grey colour.  Next on top of the two parameter diagram (FIG.~\ref{fig:stability}(a)) we put the forward (black dots) and backward transition points (magenta dots) obtained from numerical simulations for a comparison. For numerical simulations we use $N=1000$ node and use the similar setup described in section \ref{sec:obser}. The numerically computed points are found to lie precisely on the analytical curve. These findings demonstrate that the analytical theory correctly captures the system's dynamics.

Moving on to the stability diagram on the $K_2$ - $\beta$ plane  for $K_1=2.5$ (FIG.~\ref{fig:stability}(b)), it is found that the region is divided into three different parts, namely, incoherent (blue), coherent (maroon) and bistable (grey). Note that for $K_1 = 2.5$, the onset of SN bifurcation point as calculated from the  condition (\ref{critical_point}) takes place  at the critical phase lag  $\beta = 0.643501$ and $K_2 = 2.5$. For $\beta$ less than this critical value, the system follows continuous path and the synchronization level is also high (seen in FIG.~\ref{fig1}(b)). As $\beta$ is increased a little more, the system exhibits bistability. The numerically computed transition points (magenta dots) shown in the figure show very close match with the analytical one.    

Finally, in FIG. \ref{fig:stability}(c) we present the stability diagram on the $K_1$ - $K_2$ plane for $\beta = 0.8$. Here also the region is parted into coherent, incoherent and bistable states as obtained in the previous case. The bistable regime starts at the intersection point of the pitchfork line and the saddle node curve and the area of this regime broadened with increasing $K_2$. From FIG.~\ref{fig:stability}(b) and (c) we can conclude that $K_2$ contributes significantly to promote bistability along with providing high level of synchronization in the system. It has been verified that for other choices of the fixed values of one of the parameters, similar stability diagrams are obtained. Therefore, in summary, it is observed that $K_2$ promotes bistability in the system. Moreover, using the results of the ROM, we can appropriately tune the parameters in such a way that $\beta$ will promote first order transition in the system and it depends on the existence of two stable fixed points (one trivial and another nontrivial) separated by a nontrivial unstable fixed point.

\section{Conclusion }
In this letter, the results of our investigation on the effect of phase-lag on the transition to synchronization in all-to-all connected networks of phase oscillators in the presence of higher order interactions (upto $2$-simplex) have been presented. For the investigation, along with the detailed performance of numerical simulations of the full system, a reduced order model has been derived based on the Ott-Antonsen anastz to understand the origin of different solutions in numerical simulations. 

The numerical simulations with Lorenzian natural frequency reveals two distinct effects of phase lag: (1) For fixed $K_2$, the system exhibits discontinuous transition to synchronization with accompanying hysteresis loop in presence of low $\beta$ including $0$ value. As $\beta$ is gradually increased, the width of the hysteresis loop gradually decreases and eventually, the system exhibits continuous transition to synchronization when $\beta$ crosses a critical value. The appearance of discontinuous transition in globally coupled phase oscillators in presence of HOI is consistent with the earlier report~\cite{skardal2020higher}. Further, the migration of the system from the discontinuous to continuous transition states with the increment of $\beta$ in this case occurs due the enhancement of the importance of pairwise coupling in the network which is typically observed in a network in the presence of only pairwise couplings~\cite{kundu2017transition, kundu2019synchronization}.  (2) Surprisingly, completely opposite scenario is observed in numerical simulations as the higher order coupling parameter ($K_2$) is varied for a fixed $K_1$.  For lower $\beta$, transition is found to be continuous and system exhibits discontinuous transition as $\beta$ is increased beyond a critical value. Therefore, contrary to negative role of phase frustration on the discontinuous transition observed in networks with pairwise coupling, using the proposed framework with HOI, discontinuous transitions can be achieved in a network with very high phase frustration.

To understand the origin of the counter intuitive numerical simulation results mentioned above, a reduced order model is derived in the thermodynamic limit which faithfully mimic the dynamics of the original system. The bifurcation analysis of the ROM reveals a complex dependence of the transition scenario on the parameters of the system. It is observed that the first order transition is always associated with a SN bifurcation. In the first case mentioned above, the SN responsible for discontinuous transition is associated with a subcritical pitchfork bifurcation, while in the second case, SN bifurcation independently occur in a region of the parameter space. Further, all the analytically derived transition points determined from the ROM show a very good agreement with the ones determined from the numerical simulation. 

\bibliographystyle{apsrev4-1}
\bibliography{References}

\end{document}